\documentclass[twoside,slac_one]{revtex4}
\usepackage{graphicx}
\usepackage{fancyhdr}
\usepackage{amsmath} 
\usepackage{bm}
\usepackage{amsxtra}
\usepackage{amssymb}
\usepackage{amsthm}
\usepackage{latexsym}
\usepackage{lscape}

\begin{document}

\title{Model Independent Search at the D0 experiment}

%

\author{P.Renkel on behalf of theD0 collaboration}
\affiliation{Department of Physics, Southern Methodist University, Dallas, TX, USA}

\begin{abstract}
We present a review of global searches at the Tevatron with D0 detector. The strategy
involves splitting the data from the Tevatron into many final states and looking for signs of new
physics in the high $p_T$ tails of various distributions using SLEUTH algorithm. We analyzed 117 D0 final
states and 5543 D0 distributions. No evidence of
new physics is found. The two discrepant final states arise from detector modeling issues.
\end{abstract}

\maketitle

\thispagestyle{fancy}


\section{Introduction}
The standard model of particle physics has been remarkably successful: all fundamental particles predicted by this model have been discovered, with the exception of the Higgs boson. Despite
its success, there are strong motivations from the theory to expect new physics at energies at or just
above the electroweak scale.
Generally, beyond standard model theories do not give precise energy and phase space regions
to search for new physics. Motivated by this, D0[1] collaboration performed a scan
over many channels to look for significant deviations from the standard model in events containing
objects of high transverse momentum.
In this searches, we widen the scope of considered final states compared to the dedicated
analyses. At the same time, a sensitivity for each final state is generally worse than one for the
dedicated analyses.

\section{Results}

\begin{figure}
\includegraphics[scale=0.4]{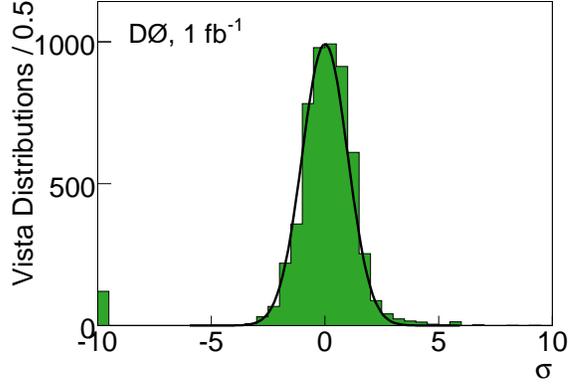}
\caption{\label{fig:sigmas}Vista histogram distribution $\sigma$ for 100\% sample before accounting for the trials factor.}
\end{figure}

 We analyze 1 fb$^{-1}$ of the Tevatron data. We generate the corresponding Monte Carlo for various processes including $Z$ and $W$ boson production, diboson and $t\bar{t}$ production. We use data to simulate the QCD processes,
and performs a fit to obtain various Scale Factors assigned to each process.
We then define exclusive final states by considering objects such as isolated electrons, muons,
taus, photons, jets, missing transverse energy and taking various combinations of those( for example electon + muon + 2 jets + missing transverse energy). We impose some transverse momentum
and pseudorapidity cuts on the selected objects. We consider only final states with at least one
lepton to avoid dealing with QCD processes that are hard to simulate.
We then perform two sorts of checks called VISTA
in the bulk of various distributions on those exclusive
final states. First, we perform a check on the number of events in each exclusive state; the good
ness of fit is calculated by Poisson probabilities. Second, we perform a shape-only analysis of
histograms within a state by calculating a Kolmogorov-Smirnov probability for the consistency of
the shape with the predicted Standard Model backgrounds. Both of these numbers require additional interpretation, because of the number of trials involved. When observing many final states
or many histograms, some disagreement is expected due to statistical fluctuations in the data. Thus
the Poisson probability used in determining event count agreement is corrected to reflect this multiple testing. The final state probabilities converted into standard deviations before adding the trials
factor correction are shown in Fig.~\ref{fig:sigmas}. We examines 117 final states and 5543 distributions and find 2 discrepant final states and 16 shape discrepancies - all of them seem to be discrepant due to difficulties in modeling of our detectors.

\begin{figure}
\includegraphics{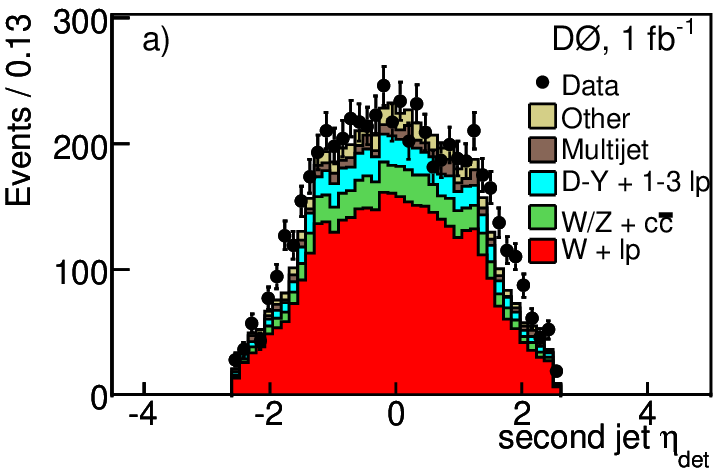}
\includegraphics{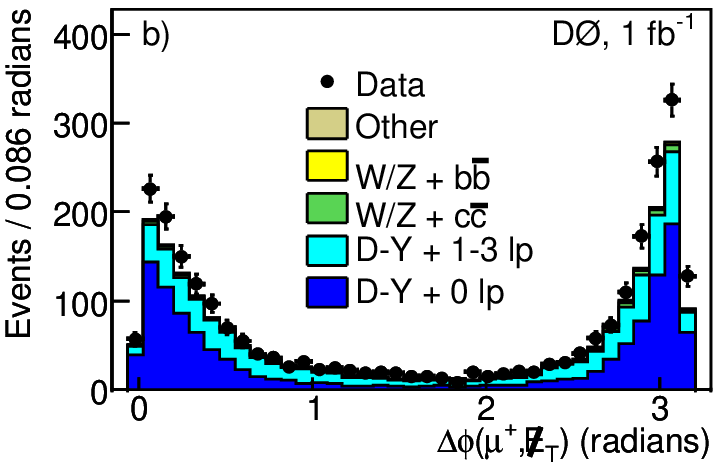}
\caption{\label{fig:modeling}The two discrepant distributions at VISTA level. The discrepancies arise due to modeling issues.}
\end{figure}

The discrepancy for the $\mu + 2 jets + \not\!\! E_T$ final state
shows the greatest difference from the SM prediction in
the modeling of jet distributions. There is a significant
excess in the number of jets at high pseudorapidities, which points to
likely problems with modeling ISR/FSR jets in the forward
region, as can be seen in Fig.~\ref{fig:modeling}a. This difference is
also observed in dedicated analyses.

The $\mu^+\mu^-$ discrepancy can be attributed to difficulties
modeling the muon momentum distribution for
high $p_T$ muons. The muon smearing
modeling is based on muons from Z and J/$\psi$ decays,
dominated by muons below 60 GeV, and is not as reliable
at high $p_T$ . The prime signature of poorly simulated high
$p_T$ muons is an excess of $\not\!\! E_T$ because of the mismodeling
of the resolution of the mismeasured track. The $\Delta \phi$ between
the positive muon and $\not\!\! E_T$ in the $\mu^+\mu^- + \not\!\! E_T$ final
state is shown in Fig.~\ref{fig:modeling}b, where the excess tends to be
for events where the $\not\!\! E_T$ is collinear with a muon.

\begin{figure}
\includegraphics{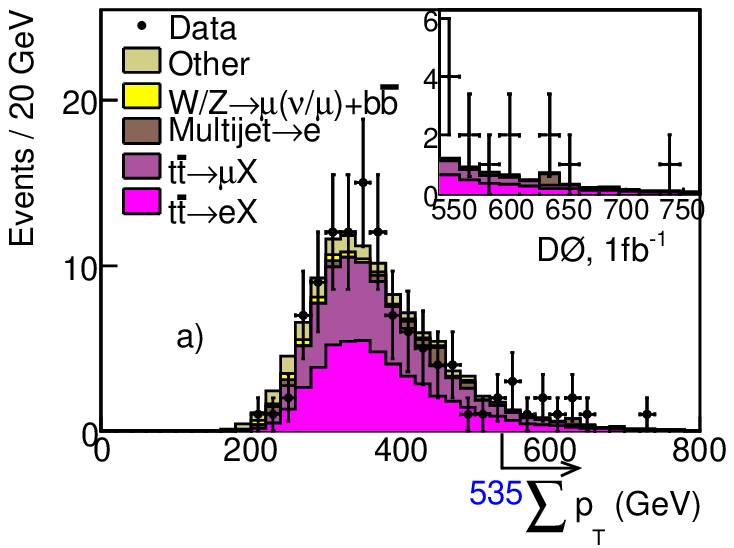}
\includegraphics{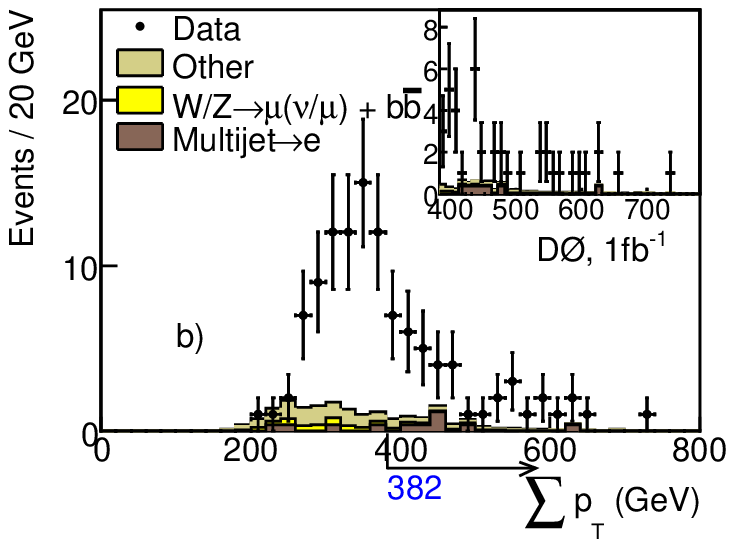}
\caption{\label{fig:ttbar}Sleuth plots with (left) and without (right) $t\bar{t}$ Monte Carlo for $b\bar{b}$ + 2 jets + met .}
\end{figure}

We then use SLEUTH algorithm in an attempt to systematically search for new physics as an
excess in the tails of high $p_T$ distributions. We use a variable that adds the absolute values of the
$p_T$ of each object in the event to the $\not\!\!\! E_T$.
We cut on the value of this variable that gives the least probability $\tilde{P}$ for the Monte Carlo to be
consistent with data. We declare the state to be discrepant after trials if this probability crosses
the threshold of $10^{-3}$. The SLEUTH algorithm is often described as being quasi-model independent,
where "quasi" refers to the assumption that the first new physics will appear as an excess of events
with high-$p_T$ objects. Thus, SLEUTH would be expected to be most sensitive to high-mass objects
decaying into relatively few final state particles.
Before we proceed, we test the SLEUTH algorithm. The question we want to answer is will
we be able to re-discover the top pairs had they not been discovered. For that, we remove the
$t\bar{t}$ processes
from our generation, and run SLEUTH. The results are presented on Fig.~\ref{fig:ttbar}. The
probability that the Monte Carlo after trial factors agrees with data is much smaller than the $10^{-3}$ -
the threshold to claim discovery.

We then run the SLEUTH on data. The most discrepant final states are given in Tab.~\ref{tab:discfs}. 
\begin{table}
\begin{tabular}{|c|c|}
\hline
Final state & p - value \\
\hline 
$l^{+}l^{-}+\not\!\! E_T$ & $<0.001$ \\
$l^{\pm}+2jet+\not\!\! E_T$ & $<0.001$ \\
$l^{\pm}+\tau^{\mp}+\not\!\! E_T$ & 0.0050 \\
$l^{\pm}+1jet+\not\!\! E_T$ & 0.019 \\
$e^{\pm}\mu^{\mp}+2b+\not\!\! E_T$ & 0.12 \\
$l^{\pm}\tau^{\mp}+2j+\not\!\! E_T$ & 0.12 \\
$l^{\pm}+2b+\not\!\! E_T$ & 0.3 \\
$e^{\pm}\mu^{\mp}+2b+\not\!\! E_T$ & 0.31 \\
$l^{\pm}+\tau^{\mp}$ & 0.91 \\
$l^{\pm}+2b+2j+\not\!\! E_T$ & 0.98 \\
\hline
\end{tabular}
\caption{\label{tab:discfs} The list of the most discrepant SLEUTH final states with the p-value after correction for trials.}
\end{table}

\begin{figure}
\includegraphics{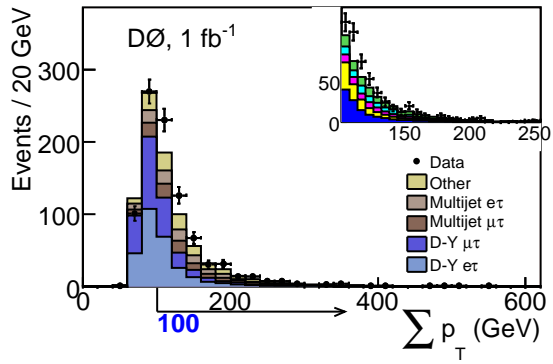}
\caption{\label{fig:dataMC}Sleuth plot in $l e^{+}\tau^{-}$ channel.}
\end{figure}

No
 final states in addition to ones discrepant at VISTA level
 surpass the discovery threshold. Fig.~\ref{fig:dataMC} shows the comparison of data and Monte
Carlo for the $l^+\tau^-$ SLEUTH final states.
\section{Conclusion}
Performing global searches with SLEUTH, we did not find any hint of new
physics in the D0 data, more data has already been recorded by the experiment.
If we incorporate this data set into our analysis and continue implementing improvements to our
correction model, we will become much more sensitive to possible new physics.

\bigskip 

\end{document}